\documentclass[aps,prc,letterpaper,11pt
,twoside,tightenlines,nofootinbib,showpacs
,preprint]{revtex4-1}
\usepackage{graphicx}
\usepackage{epsfig,float}
\usepackage{subfig}
\usepackage{amsmath}
\usepackage{amssymb}
\usepackage{amsfonts}
\usepackage{cancel}
\newcommand{\eps}{\epsilon}
\newenvironment{Tabular}[2][2]
  {\def\arraystretch{#1}\tabular{#2}}
  {\endtabular}

\newcommand{\be}{\begin{equation}}
\newcommand{\ee}{\end{equation}}
\newcommand{\ba}{\begin{eqnarray}}
\newcommand{\ea}{\end{eqnarray}}

\begin{document}
\title{The $X(3872) \to J/\psi \gamma$ decay in the $D \bar D^*$ molecular picture.}

\author{F. Aceti, R. Molina and E. Oset}
\affiliation{
Departamento de F\'{\i}sica Te\'orica and IFIC, Centro Mixto Universidad
de Valencia-CSIC,
Institutos de Investigaci\'on de Paterna, Aptdo. 22085, 46071 Valencia,
Spain}

\date{\today}

 \begin{abstract}
From a picture of the $X(3872)$ where the resonance is a bound state of $\bar{D}D^*-c.c.$, we evaluate the decay width into the $J/\psi \gamma$ channel, which is sensitive to the internal structure of this state. 

For this purpose we evaluate the loops through which the $X(3872)$ decays into its components, and the $J/\psi$ and the photon are radiated from these components. We use the local hidden gauge approach extrapolated to $SU(4)$ with a particular $SU(4)$ breaking. The radiative decay involves anomalous couplings and we obtain acceptable values which are compared to experiments and results of other calculations.

Simultaneusly, we evaluate the decay rate for the $X(3872)$ into $J/\psi \omega$ and $J/\psi \rho$, and the results obtained for the ratio of these decay widths are compatible with the experiment. 

We also show the grossly unacceptable results that come from taking only the $\bar{D}^0D^{*0}+c.c.$ component.

\end{abstract}

\maketitle

\section{Introduction}
\label{Intro}
The first observation of the $X(3872)$ decay into $J/\psi \gamma$ was reported by the BELLE collaboration in \cite{Abe:2005ix}. Later on this decay mode was confirmed by the BABAR collaboration in \cite{Aubert:2006aj} and more recently again in BELLE in \cite{Bhardwaj:2011dj}. Theoretically this decay mode had already some early attention and was studied in \cite{Barnes:2003vb,Braaten:2003he,Swanson:2004pp,Braaten:2005ai} assuming either a charmonium state or a molecular state. A thorough discussion of the different models used and the results can be seen in \cite{thomas1}, which has also been updated recently in \cite{thomas2}. A recent work assuming the $X(3872)$ to be a charmonium state is presented in \cite{Badalian:2012jz} and assuming it to be a tretaquark in \cite{Dubnicka:2011mm}. In \cite{marina} it is assumed to be a mixture of a charmonium and a molecular component, and uscing QCD sum rules a good rate is obtained for the $J/\psi\gamma$ decay mode versus the $J/\psi\pi^+\pi^-$ one, which is evaluated in \cite{navarra}. In \cite{thomas1} the authors consider, like in the work of  \cite{Braaten:2003he},  the $X(3872)$ resonance to be a molecule of $D^0 \bar D^{*0} -c.c.$ and, in addition, they include the possibility of a $c \bar c$ admixture. In \cite{thomas1} an effective Lagrangian is postulated to provide the coupling of the $X(3872)$ to the  $D^0 \bar D^{*0}$ components, with an unknown wave function. The effective coupling needed in the loops for radiative decay of the $X(3872)$ is obtained using the Weinberg compositness condition  \cite{composite,hanhart}, reformulated in \cite{sumrule} as $g^2=-(\frac{\partial}{\partial_s}G)^{-1}$, where $G$ is the loop function of the $D^0$ and  $\bar D^{*0}$ propagators. The procedure has been shown to provide a fair description of the molecular states in other works \cite{Faessler:2007gv,Branz:2008cb,Dong:2009yp}. The results of \cite{thomas1} are tied to unknowns on the regularization of the loop functions, the $\Lambda_M$ parameter used in \cite{thomas1}, and the binding. The results obtained for the $X(3872)$ decay into the  $J/\psi \gamma$ channel are of  about 125-250 KeV taking reasonable values for the $\Lambda_M$ parameter between 2 and 3 GeV. 

In \cite{thomas2} the authors include the charge components of $D^+ D^{*-}-c.c.$, which were found necessary to explain the ratio of $X(3872)$ to $J/\psi \rho$ and $J/\psi \omega$ in \cite{Gamermann_Oset,sumrule}.  The novelty with respect to the previous work of \cite{thomas1} is that the authors use a smaller $\Lambda_M$ cut off, of the order of 0.5 GeV to regularize the loop function, such that the wave function of the $D^0 \bar D^{*0} -c.c.$ is much more extended in space. The final results of the new evaluations differ quantitatively from those of \cite{thomas1} and are now in the range of 2-17 KeV. It is then clear that a more systematic approach to the problem has to be done if one wishes to obtain accurate numbers from the molecular picture of the $X(3872)$. This is the purpose of the present paper. 

A consistent dynamical picture of the $X(3872)$ in the coupled channels of $D \bar  D^* -c.c.$ was elaborated in \cite{Gamermann_Oset} using an extrapolations to SU(4) of chiral Lagrangians used in the study of pseudoscalar meson interaction with vector mesons \cite{roca}. This is equivalent to extending to $SU(4)$ the local hidden gauge approach of \cite{hidden1,hidden2,hidden3,hidden4} with a particular $SU(4)$ breaking. Given the subtlety of the small binding for the neutral $D^0 \bar  D^{*0} -c.c.$ component versus the about 7 MeV binding for the charged $D^+  D^{*-} -c.c.$ components, a coupled channel approach considering these explicit channels with their exact mass and not assuming isospin symmetry, was done in \cite{danisospin}, concluding that the coupling of the resonance to the neutral and charged components was very similar, which tells us that in strong processes the $X(3872)$ behaves as a rather good I=0 object. The $D_s ^+  D_s^{*-} -c.c.$ components were also included in \cite{Gamermann_Oset,danisospin}. On the other hand, the study of the wave functions of the resonance for each of its channels indicated that the couplings provided essentially the wave function at the origin, which indicates that the neutral component, extending further away in space than the charged components, because of the small binding, and having the same strenght at the origin, will have a larger probability in the wave function. But this has not repercussion in strong processes, of short range, or electromagnetic ones where there is production of real photons radiated from the components of the wave function, because they see the strength of the wave functions at the origin. In sum, the charged $D^+  D^{*-} -c.c.$ components have to be considered in the study of these processes.

In the present work we follow the approach of \cite{danisospin,sumrule} where all the couplings are accurately determined from the unitary coupled channel approach and are tied to the binding of the  $X(3872)$, which is generated dynamically as a composite state  of $D \bar D^*$ in this picture. The mechanisms for radiative decay are then basically the same as in \cite{thomas2}, except that we also have contribution from the $D_s \bar D^*_s$ components and have, although not much, different couplings of the resonance to the neutral and charged $D \bar  D^*$ components. The work is also technically different. Our approach has not ambiguities about the regularization of the loops, and most of the terms are shown to be convergent. Some terms are formally divergent, but we can isolate the divergence into a term proportional to the same loop function G which appears in scattering. The function G is regularized in the scattering problem in order to fit the position of the resonance, so when it comes to evaluate the radiative decay it is already fixed. This makes the scheme fully selfconsistent and predictive, since one does not have to rely upon unknown parameters that have proved to have a strong repercussion in the numerical results from the former studies.

Traditionally the $X(3872)$ could be considered as a $J^P=1^{++}$ or       $J^P=2^{-+}$ state, and there is a work similar to the one of \cite{thomas2} but assuming $J^P=2^{-+}$  \cite{Harada:2010bs}. Here we will continue to use the  $J^P=1^{++}$,  which is supported by recent analysis of data in  \cite{hanhart1++,Ke:2011jf}. 

Our work proceeds as follows: in the next section we present the formalims for the work with the Feynman diagrams used and the scheme to evaluate them. In section \ref{res} we present the results for $X(3872)\rightarrow J/\psi\gamma,J/\psi\omega,J/\psi\rho$ and compare them to experiment, discussing the role of the charged components of the $X(3872)$ wave function. In section \ref{conclusions} we summarize our results.                                  

\section{Formalism}
\label{form}
In \cite{Gamermann_Oset, x3872}, the interaction between pseudoscalars and vector mesons is studied including the charm sector. The potential is like the Weinberg-Tomozawa interaction between pseudoscalar mesons but including the vector meson fields \cite{roca}. In \cite{Gamermann_Oset, danisospin}, all the different currents within the $SU(4)$ scheme are classified in terms of $SU(3)$ currents, and the breaking symmetry parameters are introduced to account for the suppression of the heavy meson exchange. Within this formalism, the $X(3872)$ is a dynamically generated resonance from the interaction of $D\bar{D}^*$, having an eigenstate of positive $C$-parity with isospin $I=0$. It also has some component of $D_s\bar{D}_s^*$. In fact, the basis of positive $C$-parity and $I=0$ for these two channels corresponds to:
\begin{equation}
\begin{split}
\label{eq:states}
&\frac{1}{\sqrt{2}}|(D^*\bar{D}-\bar{D}^*D), I=0, I_3=0\rangle=\frac{1}{2}|(D^{*+}D^--D^{*-}D^++D^{*0}\bar{D}^0-\bar{D}^{*0}D^0)\rangle\\
&\frac{1}{\sqrt{2}}|(D^*_s\bar{D}_s-\bar{D}^*_sD_s), I=0, I_3=0\rangle=\frac{1}{\sqrt{2}}|(D^{*+}_sD^-_s-D^{*-}_sD^+_s)\rangle\ .
\end{split}
\end{equation}
In \cite{danisospin} it was found that the $X(3872)$ had couplings to the charged and neutral components of $DD^{*}$ that were very close to each other, implying an approximate $I=0$ character for the state. Since the masses and bindings used in \cite{Gamermann_Oset, danisospin} have been updated, we have redone the calculation of \cite{Gamermann_Oset, danisospin} with updated masses, assuming the present binding of $0.2\ MeV$ of the $X(3872)$ with respect to the $D^0\bar{D}^{*0}-c.c$ component. The result of the couplings are shown in table \ref{tab:x3872}.   
\begin{table}[htpb]
\centering
\begin{Tabular}{lr}
Channel&$|g_{R\to PV}|$ [MeV]\\\hline
$(K^-K^{*+}-c.c.)/\sqrt{2}$&$-53$\\
$(K^0\bar{K}^{*0}-c.c.)/\sqrt{2}$&$-49$\\
$(D^-D^{*+}-c.c.)/\sqrt{2}$&$3638$\\
$(D^0\bar{D}^{*0}-c.c.)/\sqrt{2}$&$3663$\\
$(D^-_s D^{*+}_s-c.c.)/\sqrt{2}$&$3395$\\
\hline
\end{Tabular}
\caption{Couplings $g_R$ of the pole at $(3871.6-i0.001)$ MeV to the channels ($\alpha_H=-1.27$ here). Table taken from Ref. \cite{danisospin}.}
\label{tab:x3872}
\end{table}

From the couplings in Table \ref{tab:x3872}, we observe that there is some isospin violation, which is however very small, less than $1\%$. Intuitively, one might think that the $D^0\bar{D}^{*0}$ component is the only relevant, because the binding of the $D⁰\bar{D}^{*0}$ is very small, of the order of $0.2\ MeV$ and the wave function extends much further than for the charged component, which is bound by about $8\ MeV$. However, as we mentioned, the relevant interactions in most processes are short ranged and then the wave functions at the origin, proportional to the couplings, are what matters. In this sense it is found in \cite{Gamermann_Oset, danisospin} that in the limit of the $D^0\bar{D}^{*0}$ binding going to zero, all couplings go to zero, but the ratio of the couplings of the charged and neutral components goes to a constant close to one, which guarantees that the charged component will play an important role in physical processes. The conclusion is that the wave function of the $X(3872)$ is very close to the isospin $I=0$ combination of $D^0\bar{D}^{*0} - c.c.$ and $D^+D^{*-} - c.c.$ and has a sizable fraction of the $D^+D^{*-} - c.c.$ of Eq. (\ref{eq:states}).

From this table we can also see that the couplings to the $K^-K^{*+}-c.c.$ and $K^0\bar{K}^{*0}-c.c.$ channels represent less than the $1\%$ of the contributions from the other channels (the $\pi^-\rho^+-c.c.$ has even smaller strenght). Therefore, we will treat the $X(3872)$ as if it were dynamically generated only from the last three channels in Table \ref{tab:x3872}.
\begin{figure}[htpb]
\centering
\includegraphics[scale=0.7]{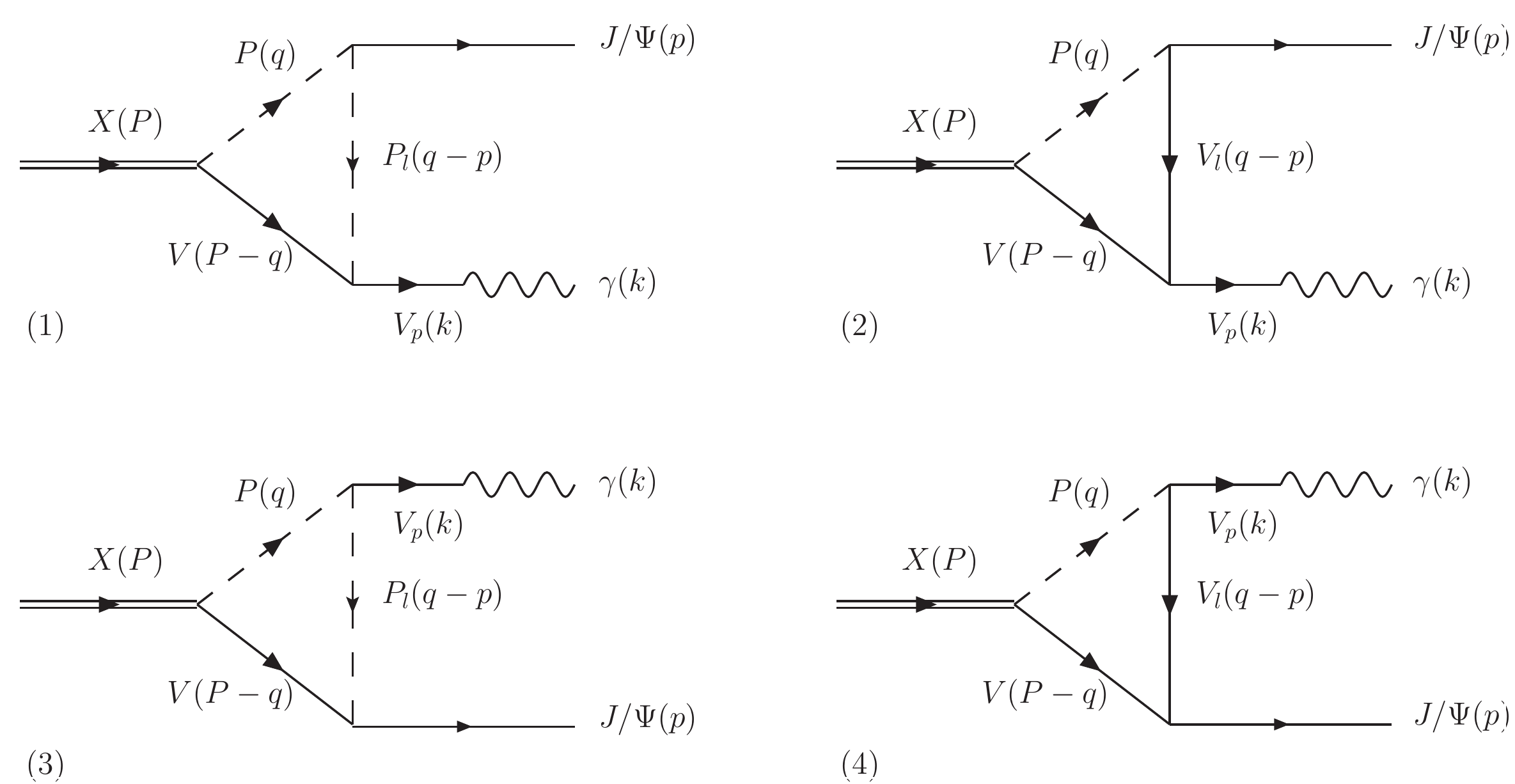}
\caption{Different types of Feynmann diagrams for the decay of the $X(3872)$ into $J/\psi\gamma$.}
\label{fig:diag}
\end{figure}

In this picture, the $X(3872)$ decays into $J/\psi \gamma$ through the diagrams shown in Fig. \ref{fig:diag}. From this figure we observe that there are  four kinds of different Feynman diagrams, all of them with an anomalous vertex coupling two vectors and a pseudoscalar (VVP),  depending on whether the diagram contains a PPV or a 3V vertex, or the photon emerges from the anomalous vertex. To begin with, there are three different channels: $D^0\bar{D}^{*0}$, $D^+D^{*-}$ and $D^+_s D^{*-}_s$, which lead to $12$, plus another $12$ for the complex congujate, Feynman diagrams to evaluate. The formalism used is very similar to the one of \cite{hideko} where the authors study the radiative decay of the dynamically generated resonance $K^{*}_2(1430)$ \cite{geng} into $K\gamma$, via diagrams containing anomalous vector-vector-pseudoscalar vertices. The VPP, 3V and V$\gamma$ vertices are evaluated using the local hidden gauge approach \cite{hidden1,hidden2,hidden3,hidden4} which automatically incorporates vector meson dominance, by means of which the photons couple to other hadrons converting itself into $\rho^0,\omega,\phi$ and $J/\psi$. As a consequence of this, we are also able to evaluate the rates of the $X(3872)$ decay into $J/\psi \rho,~J/\psi \omega$ and the ratios of the decay rates, which can be compared to existing data. 

In summary, the Lagrangians we need in order to evaluate the amplitude are listed below:
\begin{eqnarray}
&\mathcal{L}&_{VVP}=\frac{G'}{\sqrt{2}}\epsilon^{\mu\nu\alpha\beta}\langle\partial_\mu V_\nu \partial_\alpha V_\alpha P\rangle\label{lag1}\\
&\mathcal{L}&_{V\gamma}=-M_V^2\frac{e}{g}A_\mu\langle V^\mu Q\rangle\label{lag2}\\
&\mathcal{L}&_{PPV}=-ig\langle V^\mu [P,\partial_\mu P]\rangle\label{lag3}\\
&\mathcal{L}&_{3V}=ig\langle (V^\mu \partial_\nu V_\mu -\partial_\nu V_\mu V^\mu)
V^\nu)\rangle\ ,\label{lag4}
\end{eqnarray}
with $e$ the electron mass ($e^2/4\pi=\alpha$), $G' = 3g'^2/(4\pi^2f)$, $g' = -G_V M_\rho/(\sqrt{2}f^2)$, $G_V=f/\sqrt{2}$ and $g=M_V/2f$. The constant $f$ is the pion decay constant $f_{\pi}=93\ MeV$, $Q=diag(2,-1,-1,2)/3$ and $M_V$ is the mass of the vector meson, for which we take $M_{\rho}$. 

The P and V matrices contain the 15-plet of the pseudoscalars and the 15-plet of vectors respectively in the physical basis considering $\eta$, $\eta'$ mixing \cite{gamphi3770}:
\begin{equation}
\renewcommand{\tabcolsep}{1cm}
\renewcommand{\arraystretch}{2}
P=\left(
\begin{array}{cccc}
\frac{\eta}{\sqrt{3}}+\frac{\eta'}{\sqrt{6}}+\frac{\pi^0}{\sqrt{2}} & \pi^+ & K^+&\bar{D}^0\\
\pi^- &\frac{\eta}{\sqrt{3}}+\frac{\eta'}{\sqrt{6}}-\frac{\pi^0}{\sqrt{2}} & K^{0}&D^-\\
K^{-} & \bar{K}^{0} &-\frac{\eta}{\sqrt{3}}+\sqrt{\frac{2}{3}}\eta'&D^-_s\\
D^0&D^+&D^+_s&\eta_c
\end{array}
\right)\ ,
\end{equation}
and $V_\mu$ represents the vector nonet:
 \begin{equation}
\renewcommand{\tabcolsep}{1cm}
\renewcommand{\arraystretch}{2}
V_\mu=\left(
\begin{array}{cccc}
\frac{\omega+\rho^0}{\sqrt{2}} & \rho^+ & K^{*+}&\bar{D}^{*0}\\
\rho^- &\frac{\omega-\rho^0}{\sqrt{2}} & K^{*0}&D^{*-}\\
K^{*-} & \bar{K}^{*0} &\phi&D^{*-}_s\\
D^{*0}&D^{*+}&D^{*+}_s&J/\psi
\end{array}
\right)_\mu\ .
\end{equation}

From Eqs. (\ref{lag1}), (\ref{lag2}), (\ref{lag3}) and (\ref{lag4}) we can write the vertices involved in the diagram of type (1) of Fig. \ref{fig:diag} as
\begin{equation}
\label{eq:ts}
\begin{split}
&t_{RVP}=g_X\ \epsilon^{(V)\mu}\epsilon^{(X)}_{\mu}\\
&t_{V_p\gamma}=P\ M_{V_p}^2\frac{e}{g}\ \epsilon_\mu^{(\gamma)}\epsilon^{(V_p)\mu}\\
&t_{PP_lJ/\psi}=P_V\ g\ (2q-p)_\mu\eps^{(J/\psi)\mu}\\
&t_{VV_pP_l}=A\ G'\ \eps^{\alpha\beta\gamma\delta}\ (P-q)_\alpha\eps_{\beta}^{(V)}k_\gamma\eps_\delta^{(V_l)}\ ,
\end{split}
\end{equation}
where $g_X=3638/\sqrt{2}, 3663/\sqrt{2}, 3395/\sqrt{2}\ MeV$, for $D^-D^{*+}$, $\bar{D}^0D^{*0}$, $D^-_sD^{*+}$ and $-3638/\sqrt{2},$ $-3663/\sqrt{2},-3395/\sqrt{2}\ MeV$, for $D^+D^{*-}$, $D^0\bar{D}^{*0}$, $D^+_sD^{*-}_s$ respectively and $P$, $P_V$ and $A$ are numerical factors.

The $V_p\to\gamma$ conversion essentially replaces, up to a constant, $\eps_\delta^{(V_p)}$ by $\eps_\delta^{(\gamma)}$. Therefore, we can write the amplitude of the diagram (1) depicted in Fig. \ref{fig:diag} as
\begin{eqnarray}
-it_1&&=-\mathrm{B}eg_XG'\int \frac{d^4 q}{(2\pi)^4}\eps^{(V)\beta'}\eps^{(X)}_{\beta'}\eps^{(J/\psi)\mu}(2q-p)_\mu\eps^{\alpha\beta\gamma\delta}(P-q)_\alpha\eps_\beta^{(V)}k_\gamma\eps_\delta^{(\gamma)}\nonumber\\&&\times \frac{1}{q^2-m_P^2}\frac{1}{(q-p)^2-m_{P_l}^2}\frac{1}{(P-q)^2-m_V^2}\ ,
\end{eqnarray}
where $\mathrm{B}=PAP_V$ (the values of $B$ for each case are shown in Table \ref{tab:coef}). Summing over the polarizations of the internal vector, we have
\begin{equation}
\sum_\lambda \eps^{(V)}_\beta\eps^{(V)}_{\beta'}=-g_{\beta\beta'}+\frac{(P-q)_\beta(P-q)_{\beta'}}{m_V^2}\ .
\end{equation}
When contracting with the antisymmetric tensor $\eps^{\alpha\beta\gamma\delta}$, the term $(P-q)_\beta(P-q)_{\beta'}$ disappears. Thus, we have an integral like
\begin{eqnarray}
&&\int \frac{d^4q}{(2\pi)^4}\frac{ (2q-p)_\mu (p+k-q)_\alpha}{(q^2-m_P^2+i\eps)((q-p)^2-m_{P_l}^2+i\eps)((p+k-q)^2-m_V^2+i\eps)}\nonumber\\&&=
i(ag_{\mu\alpha}+bk_\mu k_\alpha+cp_\alpha k_\mu +d k_\alpha p_\mu+e p_\alpha p_\mu)\label{eq:lor}
\end{eqnarray}
because of Lorentz covariance. After contracting with the antisymmetric tensor $\eps^{\alpha\beta\gamma\delta}$ and applying the Lorentz condition $p_\mu\eps^{(J/\psi)\mu}=0$, only the coefficients $a$ and $c$ remain to be evaluated. The $a$ coefficient is related to the logarithmically divergent part of the integral in Eq. (\ref{eq:lor}) and therefore the evaluation of this coefficient needs a special  treatment as we will see later on. We arrive to an amplitude of the form
\begin{equation}
t_1=\mathrm{B}eG'g_X\eps^{\alpha\beta\gamma\delta}(a\eps_\alpha^{(J/\psi)}+cp_\alpha k\cdot\eps^{(J/\psi)})\eps^{(X)}_\beta k_\gamma\eps_\delta^{(\gamma)}\ .
\end{equation}

Now we want to evaluate  the $a$ and $c$ coefficients. We do it using the formula of the Feynman parametrization for $n=3$,
\begin{equation}
\frac{1}{\alpha\beta\gamma}=2\int^1_0 dx\int^x_0 dy\frac{1}{[\alpha+(\beta-\alpha)x+(\gamma-\beta)y]^3}\ .\label{eq:fein}
\end{equation}
In the integral of Eq. (\ref{eq:lor}), we can perform the above parametrization with 
\begin{eqnarray}
\alpha&=&(q-p)^2-m_{P_l}^2\nonumber\\
\beta&=& q^2-m_P^2\nonumber\\
\gamma&=&(p+k-q)^2-m_V^2\label{eq:abc}\ .
\end{eqnarray}
We define a new variable $q'=q+p(x-y-1)-ky$, such that the integral of Eq. (\ref{eq:lor}) can be expressed as
\begin{eqnarray}
4\int^1_0 dx\int^x_0 dy  \int \frac{d^4 q'}{(2\pi)^4} \frac{(q'+p(1-x+y)+ky)_\mu(k-q'-p(y-x)-ky)_\alpha}{(q'^2+s_1)^3}\ ,\label{eq:int2}
\end{eqnarray} 
with 
\begin{equation}
s_1=-m_{P_l}^2 +(m_{P_l}^2-m_P^2)x+(k^2+m_P^2-m^2_V)y+p^2(x-y)(1-x+y)+2pky(x-y)-k^2y^2\label{eq:s}\ .
\end{equation}
From Eq. (\ref{eq:int2}), we must take the $iag_{ \mu\alpha}$ and $icp_\alpha k_\mu$ terms. The $c$ coefficient can be evaluated very easily, since
\begin{equation}
\int \frac{d^4 q'}{(q'^2+s_1)^3}=\frac{i\pi^2}{2s_1}
\end{equation}
and we have
\begin{eqnarray}
c=\frac{1}{8\pi^2} \int^1_0 dx\int^x_0 dy \frac{y(x-y)}{s_1}\ .\label{eq:c2}
\end{eqnarray}
The evaluation of the $a$ coefficient is a little bit more elaborated. We have the identity
\begin{equation}
iag_{\mu\alpha}=-4\int^1_0dx\int^x_0dy \frac{d^4q'}{(2\pi)^4}\frac{q'_\mu q'_\alpha}{(q'^2+s_1+i\eps)^3}\ ,
\end{equation}
and after taking the trace
\begin{equation}
ia=-\int^1_0dx\int^x_0dy \frac{d^4q'}{(2\pi)^4}\frac{q'^2}{(q'^2+s_1+i\eps)^3}\ .
\end{equation}
This part is logarithmically divergent and we will relate it to the two-meson function loop $G(P)$:
\begin{equation}
G(P=p+k)=i\int\frac{d^4q}{(2\pi)^4}\frac{1}{q^2-m_P^2+i\eps}\frac{1}{(p+k-q)^2-m_V^2+i\eps}\ .
\end{equation}
Multiplying the integrand by the factor $((q-p)^2-m_{P_l}^2)/((q-p)^2-m_{P_l}^2)$ and using the Feynmann parametrization rule with the change of variable $q'=q+p(x-y-1)-ky$, we obtain:
\begin{equation}
G(P)=2i\int^1_0 dx \int^x_0 dy \int\frac{d^4 q'}{(2\pi)^4} \frac{q'^2+(ky)^2+2pky(y-x)+p^2(x-y)^2-m_{P_l}^2}{(q'+s_1)^3}
\end{equation}
and
\begin{eqnarray}
a=\frac{G(P)}{2}+\frac{1}{32\pi^2}\int^1_0dx\int^x_0 dy \frac{(ky)^2+2pky(y-x)+p^2(x-y)^2-m_{P_l}^2}{s_1+i\eps}\ .
\end{eqnarray}

Now, we want to calculate the amplitude for the second diagram in Fig. \ref{fig:diag} containing the three-vector vertex. The only difference with the previous diagram is the 3V vertex. Thus, the amplitudes corresponding to the 3V vertex and the anomalous vertex are, respectively:
\begin{equation}
\begin{split}
t_{VV_lV_p}&=V_3g\ \lbrace(q-p+k)_\mu\eps_\nu^{(V_l)}\eps^{(V)\mu}\eps^{(V_p)\nu}-(p+2k-q)_\nu\eps_\mu^{(V)}\eps^{(V_l)\nu}\eps^{(V_p)\mu}\\&+(2(p-q)+k)_\nu\eps_\mu^{(V)}\eps^{(V_l)\mu}\eps^{(V_p)\nu}\rbrace\\
t_{V_lJ/\psi P}&=AG'\ \eps^{\alpha\beta\gamma\delta}(q-p)_\alpha\eps_{\beta}^{(V_l)}p_\gamma\eps_\delta^{(J/\psi)}\ ,
\end{split}
\end{equation}
where $V_3$ and $A$ are numerical factors.

Thus, we can write the amplitude of the diagram (2) in Fig. \ref{fig:diag} as
\begin{equation}
\begin{split}
-it_2&=-eG'g_X\mathrm{C}\ \int \frac{d^4q}{(2\pi)^4}
\eps^{\alpha\beta\gamma\delta} (q-p)_\alpha\eps_\beta^{(V_l)}p_\gamma\eps_\delta^{(J/\psi)}\eps^{(X)}_{\nu'}\eps^{(V)\nu'} \\&\times
\lbrace(q-p+k)_\mu\eps_\nu^{(V_l)}\eps^{(V)\mu}\eps^{(\gamma)\nu}-(p+2k-q)_\nu\eps_\mu^{(V)}\eps^{(V_l)\nu}\eps^{(\gamma)\mu}\\&+(2(p-q)+k)_\nu\eps_\mu^{(V)}\eps^{(\gamma)\nu}\eps^{(V_l)\mu}\rbrace\frac{1}{q^2-m_P^2+i\eps}\frac{1}{(q-p)^2-m_{V_l}^2+i\eps}\nonumber\\&\times \frac{1}{(p+k-q)^2-m_V^2+i\eps}\ ,
\end{split}
\end{equation}
where $C=V_3PA$. In this process the $\bar{D}^{*0}$ is very close to be on-shell with zero three momentum. To be consistent with the approach of \cite{danisospin}, which is neglecting the three-momentum compared to the mass of the vector meson, $\vert\vec{q}\,\vert/m_V\simeq 0$,  $\eps^{(V)0}\simeq 0$, we perform the sum over polarizations as
\begin{equation}
\sum_\lambda \eps^{(V)\mu}\eps^{(V)\nu'}\simeq \delta^{(\mu\nu')_{\mathrm{spatial}}}=\delta^{ij}
\end{equation}
We also can keep the covariant formalism and remember at the end that $\mu,\nu'$ are spatial.  The way to proceed is very similar to that of the previous diagram. The second term of the $3V$ vertex proportional to $(p+2k-q)_\beta$ does not contribute, since we have $(q-p)_\alpha p_\gamma (p+2k-q)_\beta =q_\alpha(p+2k)_\beta p_\gamma\eps^{\alpha\beta\gamma\delta}$, which applying Lorentz covariance in the integral turns into a term like $(a'p_\alpha k_\beta+b'p_\beta k_\alpha)p_\gamma\eps^{\alpha\beta\gamma\delta}=0$. Therefore, we have two kinds of integrals:
\begin{eqnarray}
&&\int \frac{d^4 q}{(2\pi)^4} \frac{q_\alpha (q-p+k)_{\nu'} }{(q^2-m_P^2+i\eps)((q-p)^2-m_{V_l}^2+i\eps)((p+k-q)^2-m_V^2+i\eps)}\nonumber\\&&=i(a_1g_{\alpha\nu'}+b_1k_\alpha k_{\nu'}+c_1p_\alpha k_{\nu'}+d_1 p_{\nu'} k_\alpha+e_1 p_{\nu'} p_\alpha)
\end{eqnarray}
and 
\begin{eqnarray}
&&\int \frac{d^4 q}{(2\pi)^4} \frac{q_\alpha 2(p-q)_{\nu} }{(q^2-m_P^2+i\eps)((q-p)^2-m_{V_l}^2+i\eps)((p+k-q)^2-m_V^2+i\eps)}\nonumber\\&&=i(a_2g_{\alpha\nu}+b_2k_\alpha k_{\nu}+c_2p_\alpha k_{\nu}+d_2 k_{\alpha} p_\nu+e_2 p_{\alpha} p_\nu)\ .
\end{eqnarray}
One can see that only the coefficients proportional to $a_1, b_1, d_1, a_2 $ and $d_2$ survive. Thus, we finally get
\begin{equation}
t_2=-\mathrm{C}eG'g_X\ \eps^{\alpha\beta\gamma\delta}\lbrace (a_1\eps^{(X)}_\alpha+(b_1 k^\mu+d_1 p^\mu) \eps^{(X)}_\mu k_\alpha)\eps^{(\gamma)}_\beta+(a_2\eps_\alpha^{(\gamma)}+d_2\eps_\mu^{(\gamma)} p^\mu k_\alpha)\eps_\beta^{(X)}\rbrace p_\gamma\eps_\delta^{(J/\psi)}\ ,
\end{equation}
where now 
\begin{equation}
\begin{split}
a_1&=-\frac{G(P)}{4}-\frac{1}{64\pi^2}\int^1_0dx\int^x_0dy\frac{(ky)^2+2pky(y-x)+p^2(x-y)^2-m_{V_l}^2}{s_2+i\eps}\\
b_1&=\frac{1}{16\pi^2}\int^1_0dx\int^x_0dy\frac{y(y+1)}{s_2+i\eps}\\
d_1&=\frac{1}{16\pi^2}\int^1_0dx\int^x_0dy\frac{y(y-x)}{s_2+i\eps}\\
a_2&=-2a_1\\
d_2&=-2d_1\ ,
\end{split}
\end{equation}
with
\begin{equation}
s_2=-m_{V_l}^2+(m^2_{V_l}-m_P^2)x+(k^2-m_V^2+m_P^2)y+p^2(x-y)(1-x+y)+2kyp(x-y)-k^2y^2\ .
\end{equation}
In order to evaluate diagrams (3) and (4) in Fig. (\ref{fig:diag}), we only have to do the exchanges $k \leftrightarrow p$ and $\eps^{(\gamma)}\leftrightarrow \eps^{(J/\psi)}$ in the amplitudes of diagrams (1) and (2).

We have
\begin{equation}
t_3=\mathrm{B}eG'g^c_X\ \eps^{\alpha\beta\gamma\delta}(a\eps_\alpha^{(\gamma)}+dk_\alpha(p\cdot\eps^{(\gamma)}))\eps^{(X)}_\beta p_\gamma\eps_\delta^{(J/\psi)}\ ,
\end{equation}
with
\begin{eqnarray}
a=\frac{G(P)}{2}+\frac{1}{32\pi^2}\int^1_0dx\int^x_0 dy \frac{(py)^2+2pky(y-x)-m_{P_l}^2}{s_3+i\eps}\ 
\end{eqnarray}
and
\begin{eqnarray}
d=\frac{1}{8\pi^2} \int^1_0 dx\int^x_0 dy \frac{y(x-y)}{s_3}\ ,\label{eq:c2}
\end{eqnarray}
where
\begin{eqnarray}
s_3=-m_{P_l}^2 +(m_{P_l}^2-m_P^2)x+(p^2+m_P^2-m^2_V)y+2pky(x-y)-p^2y^2\label{eq:s}\ ,
\end{eqnarray}
for diagram (3), and
\begin{equation}
t_4=-\mathrm{C}eG'g_X\ \eps^{\alpha\beta\gamma\delta}\lbrace (a_1\eps^{(X)}_\alpha+(c_1 k^\mu+e_1 p^\mu) \eps^{(X)}_\mu p_\alpha)\eps^{(J/\psi)}_\beta+(a_2\eps_\alpha^{(J/\psi)}+c_2\eps_\mu^{(J/\psi)} k^\mu p_\alpha)\eps_\beta^{(X)}\rbrace k_\gamma\eps_\delta^{(\gamma)}\ ,
\end{equation}
with
\begin{eqnarray}
a_1&=&-\frac{G(p)}{4}-\frac{1}{64\pi^2}\int^1_0dx\int^x_0dy\frac{(py)^2+2pky(y-x)-m_{V_l}^2}{s_4+i\eps}\nonumber\\
e_1&=&\frac{1}{16\pi^2}\int^1_0dx\int^x_0dy\frac{y(y+1)}{s_4+i\eps}\nonumber\\
c_1&=&\frac{1}{16\pi^2}\int^1_0dx\int^x_0dy\frac{y(y-x)}{s_4+i\eps}\nonumber\\
a_2&=&-2a_1\nonumber\\
c_2&=&-2c_1\ ,
\end{eqnarray}
and
\begin{eqnarray}
s_4=-m_{V_l}^2+(m^2_{V_l}-m_P^2)x+(p^2-m_{V}^2+m_P^2)y+2kyp(x-y)-p^2y^2\nonumber\\
\end{eqnarray}
for diagram (4).
\begin{table}[htpb]
\centering
\subfloat{\addtolength{\tabcolsep}{-3pt}
\begin{Tabular}{lrrrr}
Diagram&P&V&\ \ \ \ $\mathrm{P_l}$&\ B\\
\hline
1&$D^0$&$\bar{D}^{*0}$&$D^{0}$&\ $\frac{4}{3\sqrt{2}}$\\
&$D^+$&$D^{*-}$&$D^{+}$&\ $\frac{1}{3\sqrt{2}}$\\
&$D^+_s$&$D^{*-}_s$&$D^{+}_s$&\ $\frac{1}{3\sqrt{2}}$\\
\hline
$\mathrm{\bar{1}}$&$\bar{D}^0$&$D^{*0}$&$\bar{D}^{0}$&\ $-\frac{4}{3\sqrt{2}}$\\
&$D^-$&$D^{*+}$&$D^{-}$&\ $-\frac{1}{3\sqrt{2}}$\\
&$D^-_s$&$D^{*+}_s$&$D^{-}_s$&\ $-\frac{1}{3\sqrt{2}}$\\
\hline
3&$D^0$&$\bar{D}^{*0}$&$D^0$&\ $0$\\
&$D^+$&$D^{*-}$&$D^+$&\ $\frac{1}{\sqrt{2}}$\\
&$D^+_s$&$D^{*-}_s$&$D^+_s$&\ $-\frac{1}{\sqrt{2}}$\\\hline
$\mathrm{\bar{3}}$&$\bar{D}^0$&\ $D^{*0}$&$\bar{D}^0$&$0$\\
&$D^-$&$D^{*+}$&$D^-$&\ $-\frac{1}{\sqrt{2}}$\\
&$D^-_s$&$D^{*+}_s$&$D^-_s$&\ $-\frac{1}{\sqrt{2}}$\\\hline
\end{Tabular}}
\hspace{1cm}\subfloat{\addtolength{\tabcolsep}{-3pt}
\begin{Tabular}{lrrrr}
Diagram &P &V &\ \ \ \ $\mathrm{V_l}$ &\ C\\
\hline
2&$D^0$&$\bar{D}^{*0}$&$D^{*0}$&\ $0$\\
&$D^+$&$D^{*-}$&$D^{*+}$&\ $-\frac{1}{\sqrt{2}}$\\
&$D^+_s$&$D^{*-}_s$&$D^{*+}_s$&\ $\frac{1}{\sqrt{2}}$\\\hline
$\mathrm{\bar{2}}$&$\bar{D}^0$&$D^{*0}$&$\bar{D}^{*0}$&\ $0$\\
&$D^-$&$D^{*+}$&$D^{*-}$&\ $\frac{1}{\sqrt{2}}$\\
&$D^-_s$&$D^{*+}_s$&$D^{*-}_s$&\ $\frac{1}{\sqrt{2}}$\\\hline
4&$D^0$&$\bar{D}^{*0}$&$D^{*0}$&\ $-\frac{4}{3\sqrt{2}}$\\
&$D^+$&$D^{*-}$&$D^{*+}$&\ $-\frac{1}{3\sqrt{2}}$\\
&$D^+_s$&$D^{*-}_s$&$D^{*+}_s$&\ $-\frac{1}{3\sqrt{2}}$\\
\hline
$\mathrm{\bar{4}}$&$\bar{D}^0$&$D^{*0}$&$\bar{D}^{*0}$&\ $\frac{4}{3\sqrt{2}}$\\
&$D^-$&$D^{*+}$&$D^{*-}$&\ $\frac{1}{3\sqrt{2}}$\\
&$D^-_s$&$D^{*+}_s$&$D^{*-}_s$&\ $\frac{1}{3\sqrt{2}}$\\
\hline
\end{Tabular}}
\caption{Coefficients B and C of the different diagrams in Fig. \ref{fig:diag}}
\label{tab:coef}
\end{table}

\begin{figure}[htpb]
\centering
\includegraphics[scale=0.7]{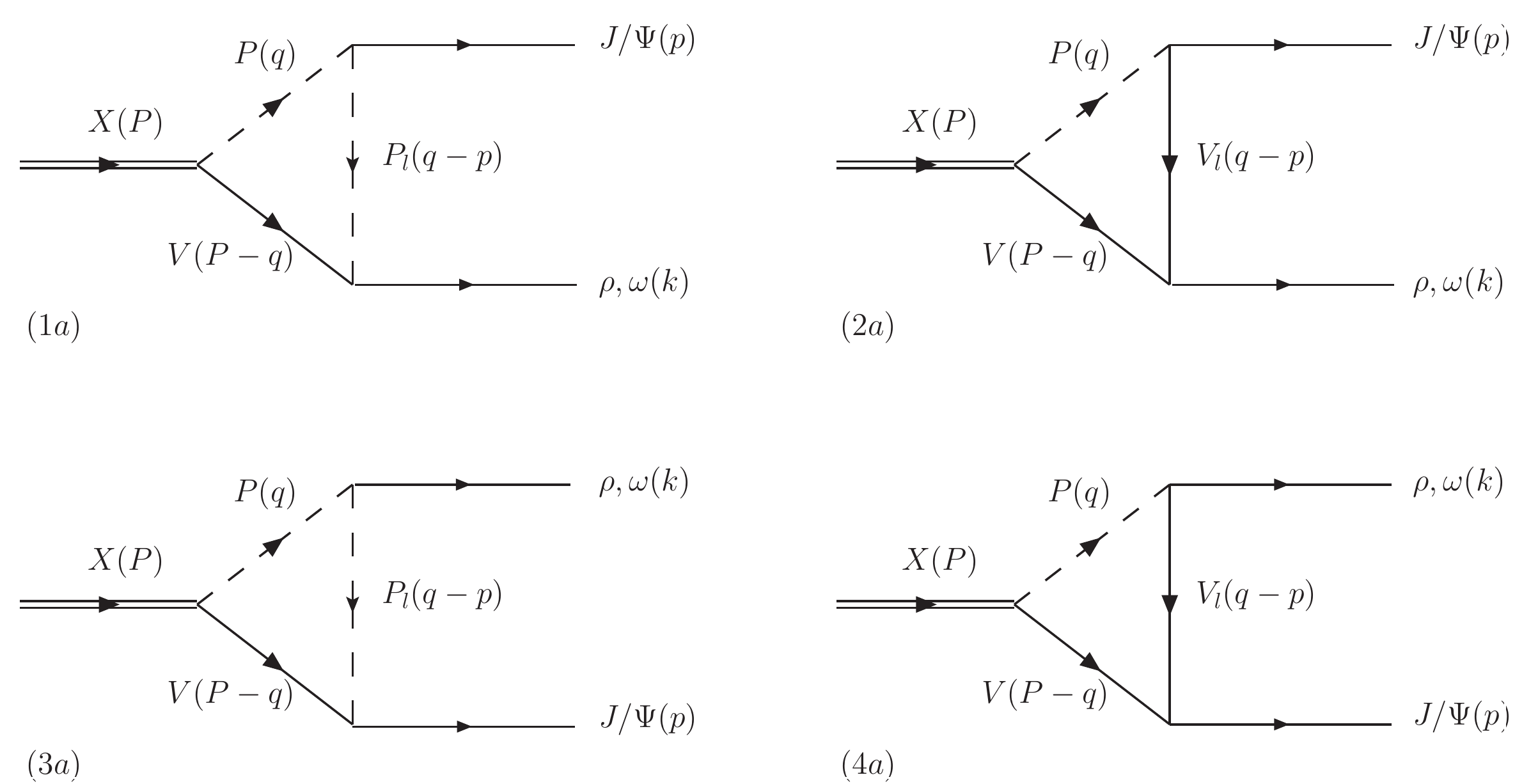}
\caption{Different types of Feynmann diagrams for the decay of the $X(3872)$ into $J/\psi\rho$ and $J/\psi\omega$.}
\label{fig:diag2}
\end{figure}

This formalism also allows us to evaluate the amplitudes for the decays $X\rightarrow J/\psi\rho$ and $X\rightarrow J/\psi\omega$ (Fig. \ref{fig:diag2}). We can proceed in complete analogy with the radiative decay to determine these amplitudes, simply removing the final photon and leaving the vector meson in the final state, the $\rho^0$ or the $\omega$. Moreover, we must take into account that the $\rho^0$ and the $\omega$ do not couple to the strange $D$ mesons, so that we have again four different kinds of diagrams, but only two channels plus their complex conjugate, that is 16 Feynman diagrams to evaluate. Doing this, we can observe that the new amplitudes have the same structure of the previous ones and can be obtained, up to a coefficient, directly with the substitutions $e\leftrightarrow g$ and $\epsilon^{(\gamma)}\leftrightarrow\epsilon^{(\rho, \omega)}$. For instance, in the case of the diagram (1a) of Fig. \ref{fig:diag2}, we have
\begin{equation}
\label{eq:trhoomega}
\begin{split}
t_{1a}=\mathrm{B'}gG'g_X\eps^{\alpha\beta\gamma\delta}(a\eps_\alpha^{(J/\psi)}+cp_\alpha k\cdot\eps^{(J/\psi)})\eps^{(X)}_\beta k_\gamma\eps_\delta^{(\rho, \omega)}\ , 
\end{split}
\end{equation}
with $a$ and $c$ the same as before
\begin{equation}
\begin{split}
a&=\frac{G(P)}{2}+\frac{1}{32\pi^2}\int^1_0dx\int^x_0 dy \frac{(ky)^2+2pky(y-x)+p^2(x-y)^2-m_{P_l}^2}{s_1+i\eps}\ ,\\ 
c&=\frac{1}{8\pi^2} \int^1_0 dx\int^x_0 dy \frac{y(x-y)}{s_1}\ 
\end{split}
\end{equation}
and $B'=P_VA$ of Eqs. (\ref{eq:ts}).
However, since we are dealing with different vertices, the new numerical coefficients, that we call $\mathrm{B'}$ and $\mathrm{C'}$, are now different and they are written in the tables \ref{tab:coef2} and \ref{tab:coef3}.

\begin{table}[htpb]
\centering
\subfloat{\addtolength{\tabcolsep}{-3pt}
\begin{Tabular}{lrrrr}
Diagram&P&V&\ \ \ \ $\mathrm{P_l}$&\ $B'$\\
\hline
1&$D^0$&$\bar{D}^{*0}$&$D^{0}$&\ $\frac{1}{2}$\\
&$D^+$&$D^{*-}$&$D^{+}$&\ $-\frac{1}{2}$\\
\hline
$\mathrm{\bar{1}}$&$\bar{D}^0$&$D^{*0}$&$\bar{D}^{0}$&\ $-\frac{1}{2}$\\
&$D^-$&$D^{*+}$&$D^{-}$&\ $\frac{1}{2}$\\
\hline
3&$D^0$&$\bar{D}^{*0}$&$D^0$&\ $-\frac{1}{2}$\\
&$D^+$&$D^{*-}$&$D^+$&\ $\frac{1}{2}$\\
\hline
$\mathrm{\bar{3}}$&$\bar{D}^0$&\ $D^{*0}$&$\bar{D}^0$&$\frac{1}{2}$\\
&$D^-$&$D^{*+}$&$D^-$&\ $-\frac{1}{2}$\\
\hline
\end{Tabular}}
\hspace{1cm}\subfloat{\addtolength{\tabcolsep}{-3pt}
\begin{Tabular}{lrrrr}
Diagram &P &V &\ \ \ \ $\mathrm{V_l}$ &\ $C'$\\
\hline
2&$D^0$&$\bar{D}^{*0}$&$D^{*0}$&\ $\frac{1}{2}$\\
&$D^+$&$D^{*-}$&$D^{*+}$&\ $-\frac{1}{2}$\\
\hline
$\mathrm{\bar{2}}$&$\bar{D}^0$&$D^{*0}$&$\bar{D}^{*0}$&\ $-\frac{1}{2}$\\
&$D^-$&$D^{*+}$&$D^{*-}$&\ $\frac{1}{2}$\\
\hline
4&$D^0$&$\bar{D}^{*0}$&$D^{*0}$&\ $\frac{1}{2}$\\
&$D^+$&$D^{*-}$&$D^{*+}$&\ $-\frac{1}{2}$\\
\hline
$\mathrm{\bar{4}}$&$\bar{D}^0$&$D^{*0}$&$\bar{D}^{*0}$&\ $-\frac{1}{2}$\\
&$D^-$&$D^{*+}$&$D^{*-}$&\ $\frac{1}{2}$\\
\hline
\end{Tabular}}
\caption{Coefficients $B'$ and $C'$ of the different diagrams in Fig. \ref{fig:diag2} in the case of a $\rho$ meson in the final state.}
\label{tab:coef2}
\end{table}

\begin{table}[htpb]
\centering
\subfloat{\addtolength{\tabcolsep}{-3pt}
\begin{Tabular}{lrrrr}
Diagram&P&V&\ \ \ \ $\mathrm{P_l}$&\ $B'$\\
\hline
1&$D^0$&$\bar{D}^{*0}$&$D^{0}$&\ $\frac{1}{2}$\\
&$D^+$&$D^{*-}$&$D^{+}$&\ $\frac{1}{2}$\\
\hline
$\mathrm{\bar{1}}$&$\bar{D}^0$&$D^{*0}$&$\bar{D}^{0}$&\ $-\frac{1}{2}$\\
&$D^-$&$D^{*+}$&$D^{-}$&\ $-\frac{1}{2}$\\
\hline
3&$D^0$&$\bar{D}^{*0}$&$D^0$&\ $-\frac{1}{2}$\\
&$D^+$&$D^{*-}$&$D^+$&\ $-\frac{1}{2}$\\
\hline
$\mathrm{\bar{3}}$&$\bar{D}^0$&\ $D^{*0}$&$\bar{D}^0$&$\frac{1}{2}$\\
&$D^-$&$D^{*+}$&$D^-$&\ $\frac{1}{2}$\\
\hline
\end{Tabular}}
\hspace{1cm}\subfloat{\addtolength{\tabcolsep}{-3pt}
\begin{Tabular}{lrrrr}
Diagram &P &V &\ \ \ \ $\mathrm{V_l}$ &\ $C'$\\
\hline
2&$D^0$&$\bar{D}^{*0}$&$D^{*0}$&\ $\frac{1}{2}$\\
&$D^+$&$D^{*-}$&$D^{*+}$&\ $\frac{1}{2}$\\
\hline
$\mathrm{\bar{2}}$&$\bar{D}^0$&$D^{*0}$&$\bar{D}^{*0}$&\ $-\frac{1}{2}$\\
&$D^-$&$D^{*+}$&$D^{*-}$&\ $-\frac{1}{2}$\\
\hline
4&$D^0$&$\bar{D}^{*0}$&$D^{*0}$&\ $\frac{1}{2}$\\
&$D^+$&$D^{*-}$&$D^{*+}$&\ $\frac{1}{2}$\\
\hline
$\mathrm{\bar{4}}$&$\bar{D}^0$&$D^{*0}$&$\bar{D}^{*0}$&\ $-\frac{1}{2}$\\
&$D^-$&$D^{*+}$&$D^{*-}$&\ $-\frac{1}{2}$\\
\hline
\end{Tabular}}
\caption{Coefficients $B'$ and $C'$ of the different diagrams in Fig. \ref{fig:diag2} in the case of a $\omega$ meson in the final state.}
\label{tab:coef3}
\end{table}

\section{Results}
\label{res}
Following the procedure in section \ref{form} we can obtain the total decay amplitude for the radiative decay of the $X$ meson and evaluate the correspondent decay width for this channel by means of the formula
\begin{equation}
\label{eq:ww}
\Gamma=\frac{|\vec{k}|^2}{8\pi M_{X}^{2}}\bar{\sum}\sum|t|^2\ ,
\end{equation}
where we sum over the polarizations of the final states and average over the $X$ meson polarizations. 

To evaluate the amplitude we use the dimensional regularization for the loop function. The subtraction constants used are $\alpha=\alpha_S=-1.27$, where the subscript $S$ identifies the strange channel, and they are chosen to fit the mass of the $X(3872)$ \cite{danisospin}.

Applying Eq. (\ref{eq:ww}),  we obtain
\begin{equation}
\Gamma(X\rightarrow J/\psi\gamma)=149.5\ keV\ .
\end{equation}

In order to make an estimation of the theoretical uncertainty on this quantity we perform a suitable variation of the parameters used to compute the total amplitude: the coupling $G'$ for the $VVP$ vertex (Eq. (\ref{lag1})), the axial-vector-pseudoscalar couplings $g_X$ for the three channels, and the two subtraction constants in the loop function, $\alpha$ and $\alpha_S$.

We allow the constant $f$, contained in $G'$, to vary, but keeping the relationship $G_V=f/\sqrt{2}$ and replacing $M_V=M_{\rho}$ by $M_{D^*}$. The couplings $g_X$ for the neutral and strange channels are also varied, independently, by $10\%$. On the other hand, the variation of the coupling for the charged channel is such that the ratio between it and the one for the neutral channel is kept constant, in order to preserve the isospin of the $X(3872)$. 
Then, we let the subtraction constants $\alpha$ and $\alpha_S$ vary between $-1.60$ and $-1.27$. This range is motivated by the range chosen for $f$. Indeed, going to higher values of the constant $f$ causes a decrease of the potential in the Lippman-Schwinger equation used to evaluate the scattering amplitude which determines the position of the resonance. One would need to go to more negative values of the subtraction constants $\alpha$ and $\alpha_S$ in the loop function, which appears in the $a$ coefficients, to keep the pole representing the resonance in the same position. The range is thus chosen such as to produce an effect in the pole position similar to that induced by the change in $f$.

We obtain the result
\begin{equation}
\label{eq:gamma1}
\Gamma(X\rightarrow J/\psi\gamma)=(117\pm 40)\ keV\ .
\end{equation}

We can also evaluate the branching ratios for the decays $X\rightarrow J/\psi\rho$ and $X\rightarrow J/\psi\omega$. These two decays, if we consider the $\rho$ and the $\omega$ with fixed masses, are not allowed because of the phase space, but they can occur when their mass distributions are taken into account and they are observed in the decays $X\rightarrow J/\psi\pi\pi$ and $X\rightarrow J/\psi\pi\pi\pi$ respectively. The two and three pions states are produced in the decays of the $\rho$ and the $\omega$. 

Thus, the decay widths, convoluted with the spectral functions, are given by the formula
\begin{equation}
\label{eq:convolution}
\begin{split}
\Gamma_{\rho/\omega}= \frac{1}{N}\int^{(m_{\rho/\omega}+2\Gamma_{\rho/\omega})^2}_{(m_{\rho/\omega}-2\Gamma_{\rho/\omega})^2} d\tilde{m}^2\left(-\frac{1}{\pi}\right)\ Im\left[\frac{1}{\tilde{m}^2-m_{\rho/\omega}^2+i\tilde{\Gamma}_{\rho/\omega}\tilde{m}}\right]
\Gamma_X(\tilde{m})\theta(m_X-m_{J/\psi}-\tilde{m})\ ,
\end{split}
\end{equation}
where
\begin{equation}
\label{eq:normal}
N=\int^{(m_{\rho/\omega}+2\Gamma_{\rho/\omega})^2}_{(m_{\rho/\omega}-2\Gamma_{\rho/\omega})^2} d\tilde{m}^2\left(-\frac{1}{\pi}\right)\ Im\left[\frac{1}{\tilde{m}^2-m_{\rho/\omega}^2+i\tilde{\Gamma}_{\rho/\omega}\tilde{m}}\right]\ ,
\end{equation}
where $\Gamma_X(\tilde{m})$ given by Eq. (\ref{eq:ww}), changing $m_{\rho}$ and $m_{\omega}$ by $\tilde{m}$.
 
In Eqs. (\ref{eq:convolution}) and (\ref{eq:normal}), $m_{\rho}=775.49\ MeV$ and $m_{\omega}=782.65\ MeV$ are the masses of the mesons, $\Gamma_{\rho}=149.1\ MeV$ and $\Gamma_{\omega}=8.49\ MeV$ are the on shell widths and
\begin{equation}
\tilde{\Gamma}_{\rho/\omega}=\Gamma_{\rho/\omega}\left(\frac{\tilde{q}}{q_{\rho/\omega}}\right)^3\ ,
\end{equation}
where $\tilde{q}$ and $q_{\rho/\omega}$ are the off shell and on shell momentum of the mesons in the center of mass reference frame:
\begin{equation}
\begin{split}
&\tilde{q}=\frac{\sqrt{\tilde{m}^2-4m_{\pi}^2}}{2}\theta(\tilde{m}-2m_{\pi})\ ,\\
&q_{\rho/\omega}=\frac{\sqrt{m_{\rho/\omega}^2-4m_{\pi}^2}}{2}\ .
\end{split}
\end{equation}
We call $\Gamma_X$ the total decay width of the $X$ into $J/\psi\rho$ or $J/\psi\omega$ to simplify the notation and $m_{\pi}$ is the pion mass.

Using Eq. (\ref{eq:convolution}) we find
\begin{equation}
\label{eq:rhoomega}
\begin{split}
&\Gamma_{\rho}=821.9\ keV\ ,\\
&\Gamma_{\omega}= 1096.6\ keV\ ,\\
\end{split}
\end{equation}
and when the error anlysis that leads to Eq. (\ref{eq:gamma1}) is done, the band of values becomes
\begin{equation}
\label{eq:rhoomega2}
\begin{split}
&\Gamma_{\rho}=(645\pm 221)\ keV\ ,\\
&\Gamma_{\omega}=(861\pm 294)\ keV\ .\\
\end{split}
\end{equation}

With the results of Eq. (\ref{eq:rhoomega}) we can evaluate the ratio
\begin{equation}
\label{eq:ratio}
R=\frac{\mathcal{B}(X\rightarrow J/\psi\pi\pi\pi)}{\mathcal{B}(X\rightarrow J/\psi\pi\pi)}=\frac{\Gamma_{\omega}}{\Gamma_{\rho}}=1.33\ .
\end{equation}
However, the experiment gives the ratio
\begin{equation}
\label{eq:ratioexp}
R^{exp}=\frac{\mathcal{B}(X\rightarrow J/\psi\pi^+\pi^-\pi^0)}{\mathcal{B}(X\rightarrow J/\psi\pi^+\pi^-)}=1.0\pm 0.4\pm 0.3_{(sys)}
\end{equation}
and, to compare our result with this, we must take into account that the $\omega$ decays into $\pi^+\pi^-\pi^0$ with a branching ratio $B_{\omega,3\pi}=0.892$. 

Hence, our ratio to compare with $R^{exp}$ is 
\begin{equation}
\label{eq:ratiotheo}
R^{th}=\frac{\Gamma_{\omega}}{\Gamma_{\rho}}\times B_{\omega,3\pi}=1.19\ ,
\end{equation}
well within the experimental error.

The result we obtain for the ratio
\begin{equation}
\label{eq:ratio2}
\frac{\Gamma(X\rightarrow J/\psi\gamma)}{\Gamma(X\rightarrow J/\psi\pi\pi)}=0.18\ ,
\end{equation}
is also compatible with the two values known from the experiment $(0.14\pm 0.05)$ \cite{Abe:2005ix} and $(0.22\pm 0.06)$ \cite{Aubert:2006aj}.

We can also estimate the theoretical errors for the two ratios in Eqs. (\ref{eq:ratiotheo}) and (\ref{eq:ratio2}), by evaluating the $\gamma$, $\rho$ and $\omega$ decays with the same set of parameters, and varying these parameters in the range used to evaluate $\Gamma(X\rightarrow J/\psi\gamma)$:
\begin{equation}
\begin{split}
\label{eq:ratioerror}
&R^{th}=(0.92\pm 0.13)\\
&\frac{\Gamma(X\rightarrow J/\psi\gamma)}{\Gamma(X\rightarrow J/\psi\pi\pi)}=(0.17\pm 0.02)\ .
\end{split}
\end{equation} 

The uncertainties in the ratios are smaller than for the absolute values and they are of the order of the $15\%$.

Finally, we do another exercise removing the $D^+D^{*-}-c.c$ and $D_S^+D_S^{*-}-c.c$ and letting only the $D^0\bar{D}^{*0}-c.c$ contribution. The results that we obtain are
\begin{equation}
\begin{split}
\label{eq:widthsneutral}
&\Gamma_{\gamma}= 0.46\ keV\\
&\Gamma_{\rho}= 9104.9\ keV\\
&\Gamma_{\omega}= 368.9\ keV\\
&R^{th}= 0.04\\
&\frac{\Gamma(X\rightarrow J/\psi\gamma)}{\Gamma(X\rightarrow J/\psi\pi\pi)}=5.05\cdot 10^{-5} \ .
\end{split}
\end{equation}

As we can see, the two ratios that we have to compare with experiment grossy diverge from the experimental values and $\Gamma_{\rho}$ by itself becomes much bigger than the width of the $X(3872)$ ($\Gamma_X<1.2\ MeV$).

In table \ref{tab:comp} we compare our results with a variety of results available in the literature using different models. It would be interesting to test these models with the new information on the experimental ratios to help discriminate among them.

The ratio of $J/\psi\gamma$ to $J/\psi\pi\pi$ is also evaluated in \cite{thomas2}, where the Weinberg compositeness condition \cite{composite} is used to determine the couplings but other assumptions are made, and they find  a range of values from $0.18$ to $1.57$ depending on the model they consider. We should stress that once the $X$ is obtained in our case and, hence,  the couplings are determined, there is no freedom in our approach, which makes unique predictions, up to the small uncertainties tied to the limited range of variation of the parameters. This comparison helps gauge the value of the results obtained here.

We should note that our results are based on the central value of the masses in the PDG and a binding of the $D^0\bar{D}^{*0}-c.c$ of $B=0.2\ MeV$. However the error on the mass of the $X(3872)$ is as big as that value. This is important to note because the coupling goes as $B^{\frac{1}{4}}$ according to the Weinberg compositeness condition \cite{composite}. When in the future the binding can be more accurately determined we can also obtain more accurate values of the absolute rates. On the other hand, the values of the ratios will be essentially unaltered.

\begin{table}[ tp ]%
\begin{tabular}{c|c}
\hline %
\textbf{model} & $\boldsymbol{\Gamma\ [keV]}$ \\\hline
$c\bar{c}$ & $11$ \cite{Braaten:2003he}\\
$c\bar{c}$ & $139$ \cite{Swanson:2004pp}\\
molecule & $8$ \cite{Swanson:2004pp}\\
molecule & $125-250$ \cite{thomas1}\\
$c\bar{c}$ & $11-71$ \cite{thomas2}\\
molecule $+\ c\bar{c}$ & $2-17$ \cite{thomas2}\\
$2^{-+}$ & $1.7-2.1$ \cite{Harada:2010bs}\\
$c\bar{c}$ & $45-80$ \cite{Badalian:2012jz}\\
tetraquark & $10-20$ \cite{Dubnicka:2011mm}\\
present work & $77-158$\\\hline
\end{tabular}
\caption{Results from previous works for the decay width of the $X(3872)$ into $J/\psi\gamma$, using different models.}
\label{tab:comp}
\end{table}

\section{Conclusions}
\label{conclusions}
In this paper we have exploited the picture of the $X(3872)$ as a composite state of $D\bar{D}^* -c.c.$ dynamically generated by the interaction of the $D$ and $D^*$ states. The couplings of the state to the different $D \bar{D}^* -c.c.$ channels have been calculated before within this model and are used here. The coupling for the $D^0\bar{D}^{*0}-c.c$ is similar to the one that would be obtained using the compositeness condition of Weinberg, since the state is barely bound in the $D^0 \bar{D}^{*0}$ component, but the dynamics of the model produces also couplings for the $D^+D^{*-}-c.c$ and $D_S^+D_S^{*-}-c.c$ states.
Using an extension to SU(4) with an explictit breaking of this symmetry of the local hidden gauge approach, used before successfuly in the study of related processes, one can determine the widths of the $X(3872)$ to $J/\psi \rho$, $J/\psi \omega$ and $J/\psi \gamma$ and compare with the ratios determined experimentally in recent works. We find a very good agreement with the experimental results. The absolute numbers obtained for the 
different widths are also reasonable and their sum within errors, $(1.6\pm 0.6)\ MeV$, is compatible with the recent total X(3972) upper limit of the width, $\Gamma=1.2\ MeV$.

We have also conducted a test neglecting the charged and strange components of the wave 
function and thus having only the $D^0 \bar{D}^{0*} -c.c.$ component. We obtain ratios in 
great disagreement with experiment and an absolute value for the $X(3872)$ partial width 
into $J/\psi \rho$ which largely exceeds the experimental upper bound for the total width 
of the $X(3872)$. This exercise confirms the relevance of the charged channels and the 
approximate I=0 character of this resonance, contrary to the appealing picture of the 
$D^0 \bar{D}^{0*} -c.c.$ state, which has a larger probability to be found because it is less 
bound than the charged components. We showed that in these processes it is the wave 
function at the origin what matters, or more concretely the couplings, which are related 
to it, and not the probability.

\section*{Acknowledgments}
We would like to thank Dian-Yong Chen for useful informations.

 This work is partly supported by the DGICYT contract FIS2011-28853-C02-01, FEDER funds from the European Union, 
 the Generalitat Valenciana in the program Prometeo, 2009/090, and 
the EU Integrated Infrastructure Initiative Hadron Physics 3
Project under Grant Agreement no. 283286.

\end{document}